# A Direct Method for the Transient Stability Analysis of Transmission Switching Events

R. Owusu-Mireku, *Student Member, IEEE*, and H.-D. Chiang, *Fellow, IEEE*

*Abstract*-- In this paper, we propose an energy-based method for the transient stability analysis of a power system transmission switching event. In this method the exit point of pseudo-fault trajectory is used to determine a relevant controlling unstable equilibrium point (CUEP) for a switching event, the stability of the switching event is then assessed based on the energy margin between the computed relevant CUEP and the post-switching initial point. The effectiveness of the method is demonstrated on switching events in the structure-preserving models of a heavily loaded version of the WSCC 9-bus 3-machine system, and the base case IEEE 145-bus 50-machine system. A scheme for the detailed analysis of power system switching events is then proposed.

*Index Terms*-- Transient Stability, Transmission Line Switching, Transmission Switching, Direct Method, Energy Method, Lyapunov Method.

## I. Nomenclature

$\delta_i$  Rotor angle of machine $i$.
$\omega_i$  Speed of machine $i$.
$M_i$  Moment of inertia for machine $i$.
$D_i$  Damping of machine $i$.
$P_{m_i}$ Mechanical power of machine $i$.
$E'_{qi}$ Equivalent transient quadrature internal voltage of machine $i$.
$X'_{di}$ Equivalent transient reactance of machine $i$.
$Y_{ik}e^{j\alpha_{ik}}$ Network admittance.
$V_i$  Voltage magnitude at bus $i$.
$\theta_i$  Voltage angle at bus $i$.

## II. Introduction

IN the daily operations of power systems, transmission switching (TS) events occur as either disturbances or control actions. Automatic controls or operators change the configuration of the transmission system in response to faults or to improve voltage profiles or the transfer capability of transmission interfaces [1]. Transmission switching (TS) is the changing of the configuration of a power system transmission network. It could be in the form of a variation in the impedance of a network branch, for example, the use of FACTs devices and transformer tap changers, opening or closing a transmission line (transmission line switching), or bus splitting. In this work, the numerical simulations are performed on transmission line switching (TLS) events, but our proposed method applies to all switching events.

The use of TS for power system control dates to the 1980s [1],[2]. Since then, numerous research studies have been presented on efficient algorithms for finding optimal TS configurations and extensions to more control applications [3]. In [4]-[7], TS was used for steady state security control. In [8], the authors introduced the economic concept of transmission capacity bidding under market rules. In [8]-[10], TS is used as a power system economic tool. In all of these studies and applications of TS, the models used were purely static with no transient or dynamic constraints except for [11]. In [11], Chen et al. proposed a theoretical method for optimal TS with transient stability constraints. This trend of research suggests that there is a general belief that a static model for TS is sufficient for analyzing the stability of the post-switching system except in transient stability control applications. In [12], the authors presented numerical examples of power system cases where acceptable steady state solutions exist but the post-switching systems are unstable. Thus, the dynamic security assessment of post-switching events needs to be factored into the daily operation and planning of power systems.

The power system dynamic stability analysis is focused on whether a post-event trajectory will settle to an acceptable condition. Currently, only two dynamic security assessment tools can be used for the transient stability analysis of TS events: the conventional time domain simulation method and the energy-based closest UEP method. The time domain simulation method is currently the most robust method available for dynamic stability assessment. However, it is numerically demanding and consequently, time consuming. The energy function-based methods, direct methods, make this assessment without integrating the post-event system for power system transient stability analysis, by comparing the energy of the post-event initial state to a critical energy value. All energy function methods are based on Lyapunov function theory and use energy functions as an approximation of a Lyapunov function.

In this paper, we propose a new energy-based method for the transient stability analysis of transmission line switching events.

Robert Owusu-Mireku is with the School of Electrical and Computer Engineering, Cornell University, Ithaca, NY 14853, USA (e-mail: ro82@cornell.edu).

Hsiao-Dong Chiang is with the School of Electrical Engineering, Cornell University, Ithaca, NY 14853, USA (e-mail: hc63@cornell.edu).

## III. The Stability Region, a Model, an Energy Function, and Direct Methods

### A. The Stability Region

The stability region or region of attraction $A(x_s)$ of an asymptotically stable equilibrium point (SEP) $x_s$ of an ordinary differential equation $\dot{x} = f(x)$ is defined as:

$$A(x_s) \coloneq \{x \in R^n : \lim_{t \to \infty} \varphi(t,x) = x_s\}. \quad (1)$$

For an SEP $x_s$, if all the equilibrium points on its stability boundary are hyperbolic, the stable and unstable manifolds of the equilibrium points satisfy the transversality condition [13], and every trajectory on the stability boundary converges to an equilibrium point as $t \to \infty$, then the stability boundary $\partial A(x_s)$ of $x_s$ is defined as the union of the stable manifolds $W^s(x_i)$ of the unstable equilibrium points (UEP) $x_i$ on $\partial A(x_s)$ where $i = 1, 2 \cdots n$ and $n$ is the number of unstable equilibrium points on $\partial A(x_s)$ [13].

### B. Dynamic Model of the Power System

The mathematical representation of the power system transient stability problem due to a TS event comprises the following components:

Pre-Switching System:
$$[x(t), y(t)], (x_s^{pre}, y_s^{pre}), t < t_s \quad (2)$$

Post-Switching System:
$$\begin{aligned} \dot{x} &= f(x, y) \\ 0 &= g(x, y) \\ t &> t_s \end{aligned} \quad (3)$$

In the pre-switching system, the system is operating at a known stable state $(x_s^{pre}, y_s^{pre})$. When a transmission element or line is switched at time $t_s$, there is a structural change in the power system, leading to the post-switching system, represented mathematically by the differential algebraic equations (DAE) in (3). Representing the vector $(x, y)$ by $X$. If (3) has an asymptotically stable equilibrium point, $X_s(t)$, then the transient stability problem is whether or not a trajectory starting at the post-switching initial state denoted as $X(t)$ will converge to $X_s(t)$. Thus, is $X(t)$ in the stability region $A(X_s(t))$ of $X_s(t)$. Fig. 1(a) – 1(b) shows a pictorial illustration of a stable, Fig. 1(a), and an unstable, Fig. 1(b) post-switching system.

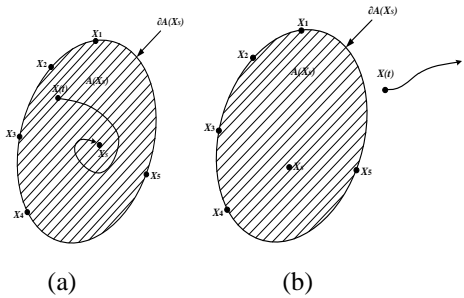

Fig.1. Depiction of a stable (a), and an unstable (b) transmission switching event

The general form of the DAE model in (3) has been thoroughly analyzed in [13]. The stability boundary of (3) has been characterized in [13] as comprising two parts: the union of the stable manifolds of an unstable equilibrium point on the stability boundary, and a collection of trajectories reaching singular surfaces. For the application of the energy based direct methods to systems of the form (3), (3) can be approximated by a two-time scale differential equation model using the singular perturbation approach (SPA) [13].

### C. An Energy Function

The function $V: R^n \to R$ is an energy function for (3) if the following conditions are satisfied:

1. Along any nontrivial trajectory $\varphi(t, x_0, y_0)$, $\dot{V}(\varphi(t, x_0, y_0)) \le 0$ and the set $\{t \in R : \dot{V}(\varphi(t, x_0, y_0)) = 0\}$ has a measure of zero in $R$.
2. If $\{V(\varphi(t, x_0, y_0)) : t \ge 0\}$ is bounded, then $\{\varphi(t, x_0, y_0) : t \ge 0\}$.

### D. Direct Methods

The direct methods for transient stability analysis are composed of the following steps:

1. Compute the initial point of the post-event system.
2. Construct an energy function for the post-event system.
3. Compute the energy function value at the post-event initial point.
4. Determine the critical point, and then compute the corresponding critical energy.
5. Compare the system energy at the post-event initial state with the critical energy. If the former is smaller, then the post-event trajectory will be stable; otherwise, it may be unstable.

The most challenging aspects of these steps are the construction of the energy function and determination of the critical point.

There is currently no analytical energy function for structure-preserving power systems with detailed generator models, controls, and large network resistance. However, numerical energy functions can be constructed for the detailed structure-preserving power system model [13]. In this work, we use the numerical energy function derived by the authors in [13].

Currently, direct methods are applied to only fault-based events or disturbances. These direct methods can be classified mainly into three groups depending on the type of critical point used in the stability assessment [13]. The first is the potential energy surface (PEBS) method. The PEBS method uses the energy at the point of maximum potential along the fault trajectory as the critical energy. The major challenge with the PEBS method is that the point of maximum potential is not always a conservative approximation of the stability boundary of the post-fault SEP [13]. The second direct method is the closest UEP method, which uses the energy at the closest UEP as the critical energy. The closest UEP is defined as the UEP on the stability boundary of the post-event SEP with the lowest energy value. The closest UEP method is known to be always conservative. The biggest challenge with the closest UEP method is the computational requirements. To find the closest UEP, we must find all the UEPs on the stability boundary of the post-event SEP, which in most cases is not an easy task. There has been much effort towards the efficient computation of the closest UEP [14]-[16], but to no avail. The third direct method for fault-based disturbance direct stability analysis is the controlling UEP (CUEP). The controlling UEP is defined as the UEP on the stability boundary of the post-event (post-fault) SEP whose stable manifold intersects with the fault trajectory. The advantage of the controlling UEP over the closest UEP method is that you only need one UEP on the stability boundary of the post-event SEP. However, finding the controlling UEP is also a challenging problem. Work in [13], [17] and others have provided a theoretical foundation and algorithmic solutions that have helped to improve the computation of the controlling UEP. In this work, we will use the BCU method presented in [13] in the computation of a CUEP in our proposed direct method for transient stability analysis of a switching event.

## IV. Direct Method for Transmission Switching Events

The following assumptions are made in this section.
1. A numerical energy function exists for the system (3).
2. The pre-switching SEP/post switching initial point is close to the post-switching SEP.
3. The energy function value at an exit point is representative of the energy function value at the corresponding CUEP. In effect, CUEPs are close to their exit points.

The PEBS and the controlling UEP methods cannot be directly applied to transient stability analysis of a switching event since these methods require a fault trajectory, which is not present in switching events. The closest UEP method, on the other hand, does not require a fault-on trajectory and hence, it is directly applicable to the transient stability analysis of a switching event. However, the closest UEP method has the problem of requiring the computation of all UEPs on the stability boundary of the post-event system, a requirement that is impractical. The energy level of the closest UEP method may also be too conservative if the post-switching initial point is not close to the portion of the stability boundary defined by the stable manifold of the closest UEP. Ideally, we need a method that can define the portion of the stability region optimally placed with respect to both the post-switching SEP and the post-switching initial point. However, such a boundary will be very difficult to define and determine.

### A. Proposed Method

We propose the use of a pseudo-fault trajectory to determine the relevant portion of the stability boundary of the post-switching SEP. The idea is to use the energy at the CUEP of a pseudo-fault applied to one of the buses of the transmission switching event as the critical energy for the direct assessment of the post-switching system stability. The proposal is as follows:
- Apply a pseudo-fault to one of the buses of the branch that is being switched.
- Determine the exit point of the sustained fault trajectory in the post-switching system using the PEBS method.
- Find the controlling UEP for the pseudo-fault trajectory in the post-switching system, using either the BCU method, the shadowing method, or others.
- Evaluate the critical energy at the computed CUEP and the energy at the post-switching initial point.
- Assess the transient stability of the post-switching system by comparing the energy at the CUEP with the energy at the post-switching initial point. If the former is greater than the latter, the post-switching system is stable; otherwise, the post-switching system may be unstable.

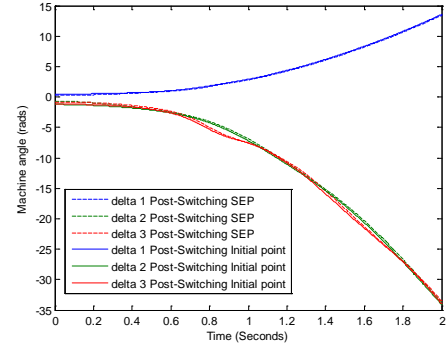

Fig. 2. Comparison of fault on trajectories, starting from the post-switching SEP and the post-switching initial point for switching contingency 1 of the WSCC 9-bus 3-machine system. The fault was applied to bus 5.

Since there are two buses at a switching branch and the CUEPs corresponding to faults at the two buses may differ, we propose the use of the bus corresponding to the CUEP with the lowest energy. Based on assumption 3, an educated estimation of the bus corresponding to the CUEP with the lowest energy can be made by comparing the energy of the two fault trajectories at their exit points. The fault trajectory with the lowest energy at the exit point will most likely have the CUEP with the lowest energy. This is because, since the exit point is in the stable manifold of the corresponding CUEP, and the minimum value of the energy function in a CUEP's stable manifold is at the CUEP itself, the smaller the energy at the exit point, the more likely it is that the energy of the CUEP will be the smallest. This choice of the fault bus will improve the conservativeness of the proposed method, and eliminate the need for checking the stability of a switching event twice.

In an unstable post-switching case, applying a fault to one of the buses of the switched branch, that starts from the projection of post-switching initial point in the fault-on system, will imply that the fault trajectory will start from outside the stability region of the post-switching system. In such a case, our proposed method will not work as expected. To overcome this challenge, we propose that the fault trajectory start from or close to the projected post-switching SEP in the fault-on system. Based on assumption 2, and since the path of fault trajectories are mostly influenced by the location of the fault, the network configuration and the unstable mode, the two fault trajectories, should be similar and close to each other, as demonstrated in Fig. 2. Fig. 2 shows a comparison of fault trajectories for a fault applied to bus 5 of the first contingency in Table I, starting from the projection of the post-switching initial point, and the projection of a post-switching SEP into the fault-on system of the WSCC 9-bus 3-machine system. We observe that the two trajectories take the same path and are practically equal.

Since applying a fault to the power system is considered much more severe than most switching events, we expect the resulting transient stability analysis results to be sufficiently conservative.

## V. SIMULATION RESULTS AND DISCUSSIONS

This study is performed on transmission line switching events using the structure-preserving models of the WSCC 9-bus 3-machine and the IEEE 145-bus 50-machine systems with classical generators. A constant impedance load model is assumed in our simulations. The generalized list of equations for the structure-preserving model is shown below.

Structure-Preserving Model: For $n$ generators and $m$ buses,

$$\dot{\tilde{\delta}}_i = \tilde{\omega}_i \quad (19)$$

$$M_i \dot{\tilde{\omega}}_i = -D_i \tilde{\omega}_i + P_{m_i} - \frac{E'_{qi} V_i \sin(\tilde{\delta}_i - \tilde{\theta}_i)}{X'_{di}} - \frac{M_i}{M_i} P_{COI} \quad (20)$$

For generator buses $i = 1, \ldots, n$:

$$(I_{di} + jI_{qi}) e^{-j(\delta_i - \pi/2)} = \sum_{k=1}^{m} Y_{ik} e^{j\alpha_{ik}} V_k e^{j\tilde{\theta}_k} \quad (21)$$

$$I_{di} = \frac{E'_{qi} - V_i \cos(\tilde{\delta}_i - \tilde{\theta}_i)}{X'_{di}}, \quad I_{qi} = \frac{V_i \sin(\tilde{\delta}_i - \tilde{\theta}_i)}{X'_{qi}}$$

For load buses $i = n+1, \ldots, m$:

$$0 = \sum_{k=1}^{m} Y_{ik} e^{j\alpha_{ik}} V_k e^{j\tilde{\theta}_k}, \quad (22)$$

$$\delta_0 = \frac{1}{M_T} \sum_{i=1}^{n} M_i \delta_i, \quad \omega_0 = \frac{1}{M_T} \sum_{i=1}^{n} M_i \omega_i,$$

$$M_T = \sum_{i=1}^{n} M_i, \quad \tilde{\delta}_i = \delta_i - \delta_0, \quad \tilde{\omega}_i = \omega_i - \omega_0,$$

$$\tilde{\theta}_i = \theta_i - \theta_0 \text{ for } i = 1, \ldots, n,$$

$$P_{COI} = \sum_{i=1}^{n} P_{m_i} - \sum_{i=1}^{n} \frac{E'_{qi} V_i \sin(\tilde{\delta}_i - \tilde{\theta}_i)}{X'_{di}}.$$

TABLE I
LIST OF LINE SWITCHING CONTINGENCIES FOR THE WSCC 3-MACHINE 9-BUS SYSTEM

| Contingency Number | From Bus | To Bus |
|---|---|---|
| 1 | 7 | 5 |
| 2 | 8 | 7 |
| 3 | 4 | 6 |
| 4 | 6 | 9 |
| 5 | 9 | 8 |
| 6 | 5 | 4 |

We computed our exit point using the PEBS method and implemented the BCU method for our controlling UEP computations [13]. The results from the proposed method are compared to the time domain simulation results and results from a brute-force implementation of the closest UEP method.

### A. Numerical Example for the WSCC 9-bus 3-machine system

In this subsection, the method is tested on the structure-preserving model of the WSCC 9-bus 3-machine system. The loading condition is set to 279.5MW real power at each load bus, with the reactive power demand kept at the same values as the base case. The test was performed on the 6 contingencies shown in Table I.

From Table II, we observe that the proposed method can detect all the unstable switching events accurately. We also observe that the closest UEP method can detect all the unstable and stable switching events accurately. For switching contingency 2, we observe that the proposed method did not produce a stability assessment result. This is because the BCU method failed to find the CUEP. In such cases, it is recommended that the contingency be tested further by detailed time domain simulation. For contingency 6, we do not have an energy margin because we were not able to compute the post-switching SEP, starting from the post-switching initial point. Such contingencies will also require further detailed analysis. For contingencies 1, 3, and 4, we observe that the closest UEP method and the proposed method have the same energy margins: the difference in energy between the critical point and the post-switching initial point. This is because the computed CUEP is the same as the closest UEP. However, for contingency 5, we observe that the closest UEP has a smaller energy margin compared to our proposed method, making the closest UEP method more conservative.

TABLE II
STABILITY RESULTS FOR THE WSCC 9-BUS 3-MACHINE SYSTEM

| Contingency # | Time Domain Simulation Stability | Closest UEP Method | | Proposed Method | |
|---|---|---|---|---|---|
| | | Stability | Energy Margin | Stability | Energy Margin |
| 1 | Unstable | Unstable | -0.0579 | Unstable | -0.0579 |
| 2 | Stable | Stable | 0.3856 | - | - |
| 3 | Unstable | Unstable | -0.3121 | Unstable | -0.3121 |
| 4 | Stable | Stable | 0.3271 | Stable | 0.3271 |
| 5 | Stable | Stable | 2.2868 | Stable | 2.3658 |
| 6 | Unstable | Unstable | - | Unstable | - |

TABLE III
LIST OF LINE SWITCHING CONTINGENCIES FOR THE IEEE 145-BUS 50-MACHINE SYSTEM

| Contingency Number | From Bus | To Bus |
|---|---|---|
| 1 | 7 | 6 |
| 2 | 14 | 17 |
| 3 | 59 | 72 |
| 4 | 115 | 116 |
| 5 | 100 | 72 |
| 6 | 91 | 75 |
| 7 | 112 | 69 |
| 8 | 101 | 73 |
| 9 | 137 | 145 |
| 10 | 139 | 145 |

TABLE IV
STABILITY RESULTS FOR THE IEEE 145-BUS 50-MACHINE SYSTEM

| Contingency # | Time Domain Simulation Stability | Proposed Method Stability |
|---|---|---|
| 1 | Stable | Stable |
| 2 | Unstable | Unstable |
| 3 | Stable | Stable |
| 4 | Stable | Stable |
| 5 | Stable | Stable |
| 6 | Stable | Stable |
| 7 | Stable | Stable |
| 8 | Stable | Stable |
| 9 | Unstable | Unstable |
| 10 | Unstable | Unstable |

### B. Numerical Example for the IEEE 145-bus 50-machine system

In this subsection, the method is tested on the structure-preserving model of the IEEE 145-bus 50-machine system. The test was performed on the 10 contingencies presented in Table III, and the stability results from the proposed method are compared to the time domain simulation stability results.

From Table IV, we observe that the proposed method detected all the unstable and stable switching events with 100% accuracy.

*C. Discussion*

The simulation results show that, despite the 3 major assumptions made in the implementation of the proposed method, the results obtained were accurate in both test systems when the underlying controlling UEP method works. It should be noted that, in some cases, the exit point of a fault trajectory cannot be computed due to the fault trajectory hitting a singular surface [18]. This is a typical challenge for the energy-based direct methods that use sustained fault trajectories. These challenges were observed for some of the fault trajectories in the simulation on the heavily loaded WSCC 9-bus 3-machine system. However, in all those instances, only one out of the two fault trajectories hit a singular surface. If the fault trajectory for both buses of a switched line hits a singular surface, the switching contingency should be evaluated with detailed time domain simulation.

As in most applications of direct methods for transient stability analysis, we recommend that our proposed method be used as a screening tool, after which the unstable switching events are sent to the time domain for detailed analysis. We therefore propose these steps for the transient stability analysis of transmission switching events.

1. Starting from the post-switching initial point, compute the post-switching SEP using the Newton method or any other fast algebraic solver. If the SEP computation fails, then the post-switching system may be unstable move to step 8; otherwise, continue to step 2.
2. Compute the energy at the post-switching initial point.
3. Starting from the projected post-switching SEP, determine the exit point of sustained fault trajectories for faults applied to the buses of the switched branch in the post-switching system, using the PEBS method.
4. Compute the energy at the two exit points and compare them to the energy at the post-switching initial point. If any one of them is less than the energy at the initial point, the post-switching system may be unstable. Skip to step 8 for time domain simulation.
5. Find the CUEP for the exit point with the lowest energy in the post-switching system using the BCU method, the shadowing method, or other methods.
6. Compute the critical energy at the CUEP.
7. Evaluate the transient stability of the post-switching system by comparing the energy at the CUEP with the energy at the post-switching initial point. If the former is greater than the latter, the post-switching system is stable; otherwise, the post-switching system may be unstable.
8. Evaluate the stability of the unstable cases detected in steps 1, 4, and 7 using detailed time domain simulations.

## VI. CONCLUSION

We have proposed a new direct method for the transient stability analysis of power system switching events. With numerical simulations, we have shown the performance of our proposed method. Finally, we presented a scheme for the screening and detailed analysis of the transient stability of switching events in the power system.